\documentclass[]{llncs}

\usepackage[utf8]{inputenc}
\usepackage[english]{babel}

\usepackage{indentfirst}

\usepackage{boxedminipage}
\newlength{\boxwidth}
\usepackage{float}
\restylefloat{figure}

\usepackage{verbatim}
\usepackage{fancyvrb}

\title{Optimizing Queries in a Logic-based\\Information Integration System}
\author{András G. Békés, Péter Szeredi}
\institute{
Budapest University of Technology and Economics\\
Department of Computer Science and Information Theory\\
1117 Budapest, Magyar tudósok körútja 2., Hungary\\
\email{bekesa@sch.bme.hu, szeredi@cs.bme.hu}\\
Keywords: information integration, query planning and optimizing, Prolog
}

\begin{document}
\pagestyle{plain}
\maketitle

\begin{abstract}
The SINTAGMA information integration system is an infrastructure for
accessing several different information sources together. Besides
providing a uniform interface to the information sources (databases,
web services, web sites, RDF resources, XML files), semantic
integration is also needed. Semantic integration is carried out by
providing a high-level model and the mappings to the models of the
sources. When executing a query of the high level model, a query is
transformed to a low-level query plan, which is a piece of Prolog code
that answers the high-level query. This transformation is done in two
phases. First, the Query Planner produces a plan as a logic
formula expressing the low-level query. Next, the Query Optimizer
transforms this formula to executable Prolog code and optimizes it
according to structural and statistical information about the
information sources.

This article discusses the main ideas of the optimization algorithm
and its implementation. 
\end{abstract}

\section{Introduction}
Integration of heterogeneous information sources requires building an
infrastructure for accessing several different information sources
together. One task of the integration is to provide a uniform
interface to the different information sources (databases, directory
servers, web services, web sites, XML files). The other task is the
semantic integration, as the meaning of the stored data can also be different
in the different sources.

In the SINTAGMA system, successor of the SILK \cite{silk:iclp} system,
semantic integration is carried out by building a high-level model and
the mappings between the high-level model and the models of the
information sources. When executing a query of the high-level model,
the query has to be transformed to queries of the sources and to the
code performing the semantic transformation of the data.  The
component of the SINTAGMA system responsible for planning and
executing queries is the Mediator. The subcomponents of the Mediator, which
translate a high-level query to a low level query are the Query
Planner and the Query Optimizer.

The output of the Query Planner is a Prolog predicate body.
This Prolog code requires call reordering to be executable: while some
sources can be called with arbitrary argument instantiations
(e.g.~predicates representing SQL tables), some predicates only can be called
when a certain subset of their arguments is instantiated (predicates
representing web services, etc).

The main and compulsory goal of the optimization step is to make the
query executable. The available modes of the predicates are given, and
with this information at hand it is decidable whether a sequence of
goals is actually callable. The secondary goal of the optimizer is to
lower the total cost of calling the query, which is basically the
estimated execution time of it. For this, some statistical information
is available on the average execution time of the predicates, and also
on the number of their solutions.

With this information at hand, the optimizer not just rearranges the
order of goals in the disjunctive branches, but it does other
manipulations on the code in the hope of obtaining a piece of code
(i.e. a query) with better performance characteristics.

Compared to the SILK system, one of the main features of SINTAGMA is
the Query Optimizer. SILK did not have this component, and query
planning usually needed manual tuning of the resulting plan. Other
important new feature of SINTAGMA is the ability to use negation and
aggregation in the queries and in the model mappings. These new
features are presented for the first time in this article.

Section \ref{sec:preliminaries} introduces the basic concepts used by
the Query Planner and Optimizer. Section \ref{sec:optimizer} introduces the main ideas of
optimizing and discusses the optimizer algorithm. Section \ref{sec:eval} discusses the
execution time issues of the implementation. Section
\ref{sec:related} compares the Query Optimizer of SINTAGMA to other systems, and Section
\ref{sec:conclusion} concludes the paper.


\section{Preliminaries}
\label{sec:preliminaries}

In the Query Planner and Optimizer, the queries are built from
\emph{predicates} using the symbols of \emph{conjunction},
\emph{disjunction}, \emph{negation} and \emph{aggregation}.

\subsection{Predicates}

Predicates, like in Prolog, represent a relation among the
\emph{arguments} of the predicate. If a predicate is \emph{called} with a
subset of its arguments \emph{instantiated}, the predicate tells the
values of the remaining arguments (by instantiating them), which
satisfy the relation. A tuple of values in relation is
called a \emph{solution} of the predicate. When a predicate is called,
it can answer with one or more solutions, or with no solutions
(\emph{failure}). Note that, in contrast with Prolog, the predicates
of our framework always instantiate all their arguments,
and an argument is either ground or uninstantiated, there are no
partially instantiated terms.

We distinguish between two kinds of predicates:

\begin{description}

\item[Source predicates:] These predicates represent
the data sources, e.g.~tables of relational databases, or methods of
web services. Arguments of the relation correspond the columns
of a database table, or the (input and output) arguments of a web service
method, etc.


\item[Constraint predicates:] These predicates represent relations among their
arguments, which are described by a known algorithm. Such predicates
are usually implemented in Prolog. When such a predicate is called,
the Mediator does not call some external entity, but answers the
predicate call by executing the algorithm. Note that these constraints
are not necessarily constraints of a CLP constraint system, they are
just Prolog predicates satisfying special requirements, as discussed later.

\end{description}

\subsection{Query Plans}

The output of the Query Planner is a query described by
the following grammar:

\begin{center}
\begin{small}
\begin{Verbatim}[numbers=none,numbersep=2pt,frame=single]
Query ::= Query,Query
       | Query;Query
       | not(Query)
       | aggregate(GroupVariables,SetExpressions,Query)
       | SourcePredicate
       | ConstraintPredicate
\end{Verbatim}
\end{small}
\end{center}
\label{fig:grammardesc}

Queries use a notation similar to that of Prolog: comma
(\verb!,!) denotes conjunction, semicolon (\verb!;!)  disjunction and
\verb!not! negation. Aggregation is explained in Section \ref{sec:aggregation}.

The result of executing a query is a set of solutions, and a solution
is a mapping, which assigns values to the variables of the query.

\subsection{I/O modes of constraint and source predicates}

Just like in Prolog, some of our predicates require a subset of their
arguments to be instantiated at the time of their call. The Mediator has to know the
allowed I/O modes of the predicates. The Query Optimizer makes queries
which respect the I/O modes of the involved predicates, and rejects a
query, if no such query plan can be made.

The I/O mode of a predicate is a mapping, which assigns
\verb!in! or \verb!out!\ to the argument positions of a predicate.
The meaning of these is similar to the
modes of the same name in Mercury, a purely declarative logic
programming language \cite{somogyi96execution}. If a mode has an
\verb!in! for an argument position, it means that the argument must be
ground when calling the predicate, an \verb!out! means that the
argument might be uninstantiated. A predicate can have several
modes. If a predicate has more than one mode, it means that when
calling the predicate, its arguments must be compatible with at least
one of its modes. If a predicate has only one mode, we often use
\emph{input} and \emph{output} as adjectives for describing arguments.

Constraint predicates behave differently than source
predicates. Constraint predicates also have I/O modes like source
predicates, but while source predicates have to be called with arguments
compatible with their modes, constraint predicates can be called any
time, independently of the instantiation of their arguments. When a 
constraint predicate is called, it checks its arguments and depending
on their state, it does the following:

\begin{enumerate}
\item If the state of the arguments of the predicate is compatible
with one of its modes, it instantiates its uninstantiated arguments
and finishes its operation.

\item If not, it ``falls asleep'', and lets the query plan
continue running. While letting the query plan run, it waits for its
arguments to be instantiated. When this happens, it goes back to step 1.

\end{enumerate}

A constraint predicate always finishes its operation when the state of
its arguments becomes compatible with one of its modes, but in certain
cases (depending on the particular predicate), it can quit
earlier. For example, the predicates representing the relations $A\geq B$
and $A\leq B$ finish their operation only when both $A$ and $B$ become
instantiated, but in the presence of each other, they can finish their
operation when one of them gets known\footnote{If these constraints
are implemented with CLP(R), the two daemons associated with the constraints quit immediately (after
unifying the variables), but for the Query Optimizer it is the
instantiation state of the constraint arguments which is important,
not the presence of daemons, and the variables are instantiated only after
one of them gets known.}.

A constraint predicate instantiates its uninstantiated arguments when
it finishes its operation, but is allowed to instantiate some of its
arguments earlier. There is no currently implemented constraint predicate
which would behave like this, but it is worth mentioning that this behaviour
is supported by the Query Optimizer.

\subsection{Optional input arguments}


When creating well-moded query plans, the Query Optimizer has to assume that when
source predicates are called and when constraints complete, they
instantiate all their uninstantiated arguments. During the development
of the Mediator, there was an increasing demand for a more flexible
handling of predicates, namely, for having \emph{optional input
arguments}.

An optional input argument (\verb!optin! in the following) is an argument of the predicate, but not
part of the relation, rather a parameter for the relation. An argument
of a predicate is an \verb!optin! argument if the predicate does not
instantiate it when the argument is uninstantiated at the time of call.

As an example for an optional input argument, let us examine a possible
information source, which is a web-service implementing a search
engine. The source has three arguments. The first argument is input,
the source expects the words to search for in this argument. The third
argument is output, the source enumerates the addresses of those
documents that contain the given words. The second argument is
optional input, it can specify the file type (.pdf, .ps, .doc, etc),
but it is not mandatory. If it is instantiated, the source answers with
only such documents which are of the given type. If not, the source
answers with addresses of files of the default type, for example, .html.

In this example we can note that this argument is not part of the
relation, as:
\begin{itemize}
\item If not instantiated at the time of call, the predicate does not
instantiate it.
\item If not instantiated, the answers are not of all the possible file types.
\end{itemize}


When dealing with optional input arguments, we have to face a problem:
The query plan is invalid if a predicate is called with an uninstantiated
optional input argument, but the argument variable is instantiated
later. This is because the optional input argument of the predicate
becomes known, but the predicate did not operate according to its
value.

The rule that describes the correct treatment of optional input
arguments is the following: A predicate is callable if its
uninstantiated optional input arguments will not be instantiated at a later point of the query.
The query planner has to make query plans that respect this rule and
has to reject a query if no such plan can be made.

Let us examine the following example with two sources:

\begin{itemize}
\item \verb!a(in,out)! (the first argument of the source has to be
instantiated (input))
\item \verb!b(optin,out)! (the first argument is optional input)
\end{itemize}

The plan \verb!a(1,X),b(X,Y)! is well-moded, because the optional
input argument \verb!X! is instantiated at the time of calling \verb!b(X,Y)!.

The plan \verb!b(X,Y),a(Y,_)! is well-moded, because although the
optional input argument \verb!X! is not instantiated at the time of
calling \verb!b(X,Y)!, it remains uninstantiated.

The query \verb!b(X,Y),a(Y,X)! is not acceptable, because:
\begin{itemize}
\item the plan \verb!a(Y,X),b(X,Y)! is ill-moded, because \verb!a(Y,X)! is
called with \verb!Y! uninstantiated.
\item the plan \verb!b(X,Y),a(Y,X)! is ill-moded, because when calling
\verb!b(X,Y)!, \verb!X! is not instantiated, but it is instantiated
later, by the call \verb!a(Y,X)!
\end{itemize}

Regarding optional input arguments, we have to enforce the following rule: If an
argument is an optional input argument in one mode, it must be an
optional input argument in all other modes as well, and the predicate
might not instantiate that argument under any condition. This is
required, because at certain points of query planning, it must be
known whether a variable might get instantiated at a later point of
the query or not\footnote{This is also required because of negations
and aggregations, discussed later.}.

\subsection{Negation}

SINTAGMA uses the closed world assumption for handling negation.
Procedurally, this is implemented as negation by failure.

A well known problem with negation by failure is when the negation is
called before all its variables are instantiated. In our framework,
this is a problem only if an uninstantiated argument of a negated predicate
gets instantiated later. To avoid this, the Query Optimizer 
considers a negated query callable (well-moded) only if its uninstantiated variables will
not be instantiated at a later point of the query.

\subsection{Aggregation}
\label{sec:aggregation}

Aggregation is used to partition the solutions of a query (``GROUP BY'' in SQL),
and combine the solutions in each partition into a single solution.
During the design of the SINTAGMA system, an important goal was to
have a query language which is at least as expressive as the query
language SQL. The Mediator of the SINTAGMA system allows the
aggregation of queries spanning several information sources, the
use of the standard SQL set functions (count, sum, min, max,
etc\dots), and the ability to extend the system with custom set
functions.

The syntax of aggregation is the following:

\begin{center}
\begin{small}
\begin{Verbatim}[numbers=none,numbersep=2pt,frame=single]
Aggregation ::= aggregate(GroupVariables,SetExpressions,Query)

GroupVariables ::= ListOfVariables
ListOfVariables ::= [] | [Variable|ListOfVariables]

SetExpressions ::= [] | [SetExpression|SetExpressions]
SetExpression ::= Variable=SetFunctionName(Argument)
Argument ::= PrologTerm
\end{Verbatim}
\end{small}
\end{center}
\label{fig:aggregation}


Let us show this construct through an example:\\
\verb!aggregate([Department],[AvgSal=avg(Salary)],!\\
\verb!          (works_at(Employee,Department),salary(Employee,Salary)))!\\
Here, \verb!Department! is the base of grouping, and \verb!AvgSal! will be bound to the
average of the \verb!Salary! values, for each group, therefore this query
returns the list of departments and the departmental average salaries. The semantics
of aggregation is the following:

\begin{enumerate}
\item \verb!Query! is executed.

\item From the solutions of \verb!Query!, groups are formed. The basis
of grouping is \verb!GroupVariables!, which is a list of some of the
variables of \verb!Query!. The members of a group are those solutions,
for which the values of variables in \verb!GroupVariables! are the same.

\item We calculate the value of each set function in the
\verb!SetExpressions! list, for each of the groups.

\item The aggregated query has one solution for each group. A solution
is the instantiation of the variables in \verb!GroupVariables!,
and the instantiation of the variables on the left-hand side of the
\verb!=! symbols in \verb!SetExpressions!. No other variables are
instantiated and no sleeping constraints are left behind.

\end{enumerate}

Note that with aggregation, there is a similar problem as with
negation, the query plan is invalid if one of the variables in the
aggregated query is uninstantiated at the time of calling the
aggregation, but later gets instantiated.

\section{Query Optimizer}
\label{sec:optimizer}

Optimizing queries means choosing the most efficient query plan among
the well-moded ones.

\subsection{Base cases of Optimization}

This subsection summarises the main query optimization techniques used in
the Query Optimizer of SINTAGMA.

\begin{description}
\item[Reordering conjunctions:]
The execution of a conjunction means executing the first member of the
conjunction, then executing the remaining part, for each solution the
first member has. As a consequence, the order of members in a
conjunction radically affects performance. It is well known that
putting a member with a small expected number of solutions to the
first place leads to a better execution time than putting a member
with a plenty of solutions \cite{DBLP:conf/acm/WangYC93}. This
optimization technique is similar to the techniques used by
the query planners of database engines \cite{ioannidis96query},\cite{chaudhuri98overview},\cite{chaudhuri95overview}, but while
the database has exact knowledge about the tables (keys, indexes,
table sizes, number of different values in columns), we have to
make do with some statistical data.

\item[Constraints first:]
In a query plan, source predicates can be
called only in places where their arguments are sufficiently
instantiated, but constraint predicates can be called anywhere. We
know when constraint predicates are bound to instantiate all their
arguments and finish their operation, but they can instantiate some of
their arguments or fail before that point. For this reason, it can be
beneficial to call constraint predicates way before their arguments
get instantiated.

\item[Postponing disjunctions:]
The part of the query plan after a disjunction is executed as many
times as many branches the disjunction has, therefore postponing disjunctions
is profitable. On the other hand, the sub-query after a
disjunction can be moved inside the branches of the disjunction, and
can be optimized differently in the different branches, which is also
beneficial. In such cases branching on disjunctions is not postponed.

\item[Delegating constraints to sources:]
Some information sources can
understand some constraints on their own. For example, an SQL database
understands a ``smaller than a given number'' constraint. When
querying such source, the constraints on the variables of the query
should be sent to the source, in order it can filter its answers
according to the constraints, as this is cheaper than transferring all
the solutions to the Mediator and filtering them there. In practice,
the source-level query sent to the sources contains the source-level
equivalents of the sleeping demons at the time of the source call.

\item[Grouping source predicates together:]
Some sources can perform
joins on their own. If some source predicates are linked through
a common variable, and they refer to the same information source, it is
beneficial to send one compound query instead of querying the source
according to the first predicate and querying it again according to
the second for each solution of the first.

\end{description}

These optimization techniques are simple and the transformations they
suggest are not particularly difficult to implement, but the
transformations contradict each other. Deciding which ones to use in certain
situations is done by estimating the cost of the resulting queries and
choosing the most promising plan.

\subsection{Optimization: the Naive Approach}

For the Query Optimizer of SINTAGMA, the following information about
the predicates is available during planning:

\begin{itemize}
\item Allowed I/O modes for each predicate
\item Expected number of solutions of a predicate for certain I/O modes
\item Expected cost (execution time) of a predicate for certain I/O modes
\end{itemize}

The simplest way of optimizing a query plan is generating all orders
of the conjunctions, throwing away the ill-moded orders, then
calculating the estimated cost of each one, and choosing the plan with the
smallest estimated cost.

Generating all the possible orders could be done with a very simple
recursive algorithm, which generates all the permutations of the
conjunctions in the query while recursively generating all the orders
of the members of conjunctions. Generating all the possible orders and
filtering out the ill-moded ones, then calculating their cost and
choosing the best would be a very inefficient way of
optimizing. Instead of that, the query optimizer interleaves these
tasks, generates only well-moded plans, calculates their cost at the
same time, and throws away partially computed plans that are known to lead to more expensive
plans than the previously found best plan. This branch-and-bound
method of finding the best plan is still exponential in execution
time, but no polynomial-time algorithm is expected as the problem is
NP-hard. Luckily, the size of the plans the Optimizer has to handle
allows us to use a well-implemented exponential algorithm, instead of
using approximation techniques for finding near-optimal solutions.

\subsection{The Optimization Algorithm}

The optimization of queries is done by a procedure with the following input arguments:
\verb!Query!, \verb!Continuation!, \verb!InstVars!, \verb!Constraints!. The output arguments are
\verb!OptimizedQuery!, \verb!ResultVars!, \verb!Cost! and \verb!NumSol!.
The result of optimization (\verb!OptimizedQuery!) is a
conjunction which starts with the optimized \verb!Query! and continues with the
optimized \verb!Continuation!. The procedure is initially
started with the query to optimize in the \verb!Query! argument and an
empty query in \verb!Continuation!. \verb!InstVars! is the set of variables that
were already instantiated by the query parts preceding \verb!Query!
and \verb!Continuation!, and \verb!Constraints! is the set of sleeping constraints.
\verb!ResultVars! is the set of variables
which are necessarily instantiated by \verb!OptimizedQuery!. \verb!Cost! and \verb!NumSol! are
the estimated cost and number of solutions of \verb!OptimizedQuery!.

With these input and output arguments, optimization can be carried out in parallel with
checking mode correctness and calculating cost. The algorithm is
described by Prolog code fragments. These code pieces give a high level view of the algorithm.
Some details of are left out,
for example the calculation of costs and number of solutions.

The task of the procedure depends on its \verb!Query!
argument. The most difficult case is when \verb!Query! is a
conjunction. The code fragment for dealing with a conjunction is shown in Figure \ref{fig:opt_conj}.
If \verb!Query! is a conjunction or a source predicate (a
conjunction of one), it first appends \verb!Continuation! to
\verb!Query! (line 1), resulting in a conjunction of many members. Then, it
chooses all the constraint predicates to fill the first places of the resulting
query (lines 3-7). When there are no more constraint predicates, it chooses each of
the members and recursively optimizes them with the remaining
members as the \verb!Continuation!.
If the chosen member is not a source predicate,
then the optimisation is simply a recursive call to \verb!optimize! (lines 9-12).

If the chosen member is a source predicate, then the optimizer tries
to pack it together with other source predicates that can be called in
succession and refer to the same information source. It does this
grouping in all possible ways (line 14).
Source query packs will also include
a suitable subset of the sleeping constraints in order to be sent to the source as well
(line 17).

\begin{figure}[htbs]
\begin{center}
\begin{small}
\begin{Verbatim}[numbers=left,numbersep=2pt,frame=single]
concat_conjunction(Query,Continuation,WholeQuery),
(
 select_from_conjunction(Element,WholeQuery,Rest),
 is_constraint(Element),
 is_callable(Element),
->
 optimize(Element,Rest,InstVars,Constraints,OptimizedQuery,ResultVars)
;
 select_from_conjunction(Element,WholeQuery,Rest),
 not_constraint(Element),
 not_source_pred(Element),
 optimize(Element,Rest,InstVars,Constraints,OptimizedQuery,ResultVars)
;
 select_callable_source_pred_sequence(SourcePreds,WholeQuery,Rest),
 % if such cannot be selected, we fail here

 create_source_query_pack(SourcePreds,Constraints,SourceQuery),
 instantiates_variables(SourcePreds,Vars),
 union(InstVars,Vars,InstVars1),
 wake_constraints(InstVars1,Constraints,InstVars2,Constraints1),
 optimize(Rest,empty,InstVars2,Constraints1,OptRest,ResultVars),
 create_conjunction(SourceQuery,OptRest,OptimizedQuery)
)
\end{Verbatim}
\end{small}
\end{center}
\caption{Optimizing Conjunctions}
\label{fig:opt_conj}
\end{figure}

Next, let us examine the case of a disjunction (Figure \ref{fig:opt_disjunction}).
If the query to be
optimized is a disjunction, it means that it is decided that the
disjunction will be the first member of a conjunction. This, however
does not mean that the predicates inside the branches of the
disjunction will precede the predicates in the continuation. It only
means that at this point the query has to fork with a disjunction.

\begin{figure}[htbs]
\begin{center}
\begin{small}
\begin{Verbatim}[numbers=left,numbersep=2pt,frame=single]
Query=(BranchA;BranchB),

concat_conjunction(BranchA,Continuation,WholeQueryA),
concat_conjunction(BranchB,Continuation,WholeQueryB),

optimize(WholeQueryA,empty,InstVars,Constraints,OptA,InstVarsA),
optimize(WholeQueryB,empty,InstVars,Constraints,OptB,InstVarsB),

intersection(InstVarsA,InstVarsB,ResultVars),

OptimizedQuery=(OptA;OptB)
\end{Verbatim}
\end{small}
\end{center}
\caption{Optimizing Disjunctions}
\label{fig:opt_disjunction}
\end{figure}

The independent optimization of the two branches and the continuation
is beneficial, because of the following: It is possible that when the
two branches of the disjunction are optimized together with the
continuation, the optimization would prefer to order the members of the branches and
the continuation in different ways, which means that the conjunction
of the optimized disjunction and the optimized continuation is
sub-optimal. This way of optimizing disjunctions is why the
\verb!Continuation! argument is needed.

There three two more cases, the optimization of negations,
aggregations and constraint predicates. The code fragments for
negation and aggregation can be seen on Figure \ref{fig:opt_not_aggr}.
The most interesting part in these, when the uninstantiated query variables are checked
whether they might be instantiated later.
The case of constraint predicates is left to the reader.

\begin{figure}[shtb]

\begin{center}
\begin{small}
\begin{Verbatim}[numbers=left,numbersep=2pt,frame=single]
Query=not(QueryInNegation),

% collecting the variables that are not known to be instantiated
collect_variables(QueryInNegation,Vars),
subtract(Vars,InstVars,UninstVars),

% checking whether Continuation or the sleeping constraints
% might instantiate one of these
instantiates_variables(Continuation,ContVars),
disjoint(UninstVars,ContVars), 		%if not, we fail here
instantiates_variables(Constraints,ConstVars),
disjoint(UninstVars,ConstVars),		%if not, we fail here

optimize(QueryInNegation,empty,InstVars,Constraints,OptQuery,_Vars),
OptNegation=not(OptQuery),
optimize(Continuation,empty,InstVars,Constraints,OptCont,ResultVars),
create_conjunction(OptNegation,OptCont,OptimizedQuery),
\end{Verbatim}
\end{small}
\end{center}

\begin{center}
\begin{small}
\begin{Verbatim}[numbers=left,numbersep=2pt,frame=single]
Query=aggregation(GroupVars,SetExprs,QueryInAggregation)
extract_aggregated_vars(SetExprs,AggVars),

% collecting the variables that are not known to be instantiated 
% and will not be instantiated by the aggregation
collect_variables(QueryInAggregation,Vars),
subtract(Vars,InstVars,UninstVars),
subtract(UninstVars,GroupVars,FlounderVars),

% checking whether Continuation or the sleeping constraints
% might instantiate one of these
instantiates_variables(Continuation,ContVars),
disjoint(FlounderVars,ContVars), 		%if not, we fail here
instantiates_variables(Constraints,ConstVars),
disjoint(FlounderVars,ConstVars),		%if not, we fail here

optimize(QueryInAggregation,empty,InstVars,Constraints,OptQuery,_Vars),
OptAggregation=aggregation(GroupVars,SetExprs,OptQuery),

union(InstVars,GroupVars,InstVars1),
union(InstVars1,AggVars,InstVars2),
wake_constraints(InstVars2,Constraints,InstVars3,Constraints1),

optimize(Continuation,empty,InstVars3,Constraints1,OptCont,ResultVars),
create_conjunction(OptAggregation,OptCont,OptimizedQuery)
\end{Verbatim}
\end{small}
\end{center}
\caption{Optimizing Negated and Aggregated Queries}
\label{fig:opt_not_aggr}
\end{figure}

The algorithm described above generates all the possible goal orders
and can simultaneously calculate their estimated costs. Note that the code
has choice-points only when dealing with conjunctions.
The \verb!optimizer! procedure succeeds at most once, resulting in
the best (cheapest) plan, or a failure if no well-moded plan can be found.

\subsection{Cost estimation}

The cost estimation of the optimizer is a field of further
research. The exact parameters of the present algorithm will be
refined during the use of the query planner in production systems.

The present algorithm implements the following ideas:

\begin{description}

\item[Constraints:]
None of the currently used constraints of the SINTAGMA system have more
than one solution. Some of the constraints either succeed or fail, but
most of them  implement functions, which means that the number of solutions is one.
The cost and the number of solutions of constraint predicates are pre-defined
constants for each of the predicates.

\item[Negation:]
The cost and number of solutions of a negated query are calculated
from the cost and number of solutions of the query by a formula which
is not fixed yet. The cost is smaller than the cost of the query, as
the query has to supply only the first solution, not all. The number
of solutions of a negated query is somewhere between 0 and 1 (it
either fails or succeeds once, with some probability).

\item[Aggregation:]
The cost of an aggregation is the cost of the aggregated query,
plus some cost of collecting the solutions for the aggregation. The
number of solutions (the number of groups) can be approximated by the
ratio of the number of solutions of the aggregated query, and the
number of solutions of it assuming that the \verb!GroupVariables! are also instantiated.

\item[Disjunction:]
The cost and number of solutions of a disjunction is the sum of
costs and number of solutions' of the two branches.

\item[Conjunction:]
The number of solutions of a conjunction is the number of
solutions of the first member, multiplied by the (recursively
calculated) number of solutions of the remaining part of the
conjunction. The cost of a conjunction is the cost of the first
member, plus the cost of the (recursively calculated) cost of the
remaining part of the conjunction multiplied by the number of
solutions of the first member.

\item[Source predicates:]
The cost and the number of solutions of source predicates are
derived from statistical data. The Mediator continuously collects
statistical data about the execution time and number of solutions of
source predicates for each instantiation state of their arguments. From
this data the query optimizer can estimate the cost and number of
solutions of a source predicate in a query plan, if the statistical
database contains information for the predicate with the same
instantiation state of its arguments. If not, the query optimizer
interpolates from the statistical data of other
instantiation states, and may also use other information. For
example, some relational information sources can tell the number of rows in a
table and the number of different values in a column of a table
(from which one can tell if a column is a key).

\item [Source query packs:]
The number of solutions of a source query pack is the same as
calling the source predicates in a conjunction. The formula to
calculate the cost of a source query pack is subject to further
research. There are many circumstances to consider: the filtering ability
of the sources
\begin{itemize}
\item if some arguments are bound to constants (this is covered by the
statistical data)
\item if some arguments of different predicates are bound to each other
\item if some arguments of a single predicate appear in a constraint
\item if some arguments of different predicates appear in a constraint
\end{itemize}

\end{description}

\section{Evaluation}
\label{sec:eval}

Although the Mediator memorizes the planned (and optimized) queries,
query planning time does matter, and extreme planning times are not acceptable.
The runtime of the above described algorithm is exponential in the
size of the input. The proposed branch-and-bound technique dramatically speeds
up the optimization code, but is still slow in some cases. However, there is an
untapped opportunity to further reduce the runtime of the computation: memoizing
the results of optimizing the sub-queries.

Let us examine how the optimizer traverses the space of possible query plans of the
query \verb!(a,b,c,...)!. First it chooses \verb!a! as the first goal of the query, and
recursively optimizes \verb!(b,c,...)!, which involves the recursive optimisation
of \verb!(c,...)!. Next, it chooses \verb!b! as the first goal of the query, and
recursively optimizes \verb!(a,c,...)!, which involves the recursive optimisation
of \verb!(c,...)!, and so on. Calculating of the best plan of \verb!(c,...)! is done
several times. This duplicated work can be eliminated, if the optimizer memoizes 
the results of optimizing sub-queries.

Memoizing is implemented as a meta-predicate that memoizes the best solution of a goal
(according to an arithmetic expression), and succeeds at most once, unifying the goal with its
best solution. It also memoizes the result if it is a failure or an exception. The meta-predicate
also gives the called goal the opportunity to read the value of its best previous result, so
it can stop traversing branches of its search space where no better solution can be found.

The Query Optimizer uses the memoizer for all of its recursive calls, plus reads the
value of its best previous result, and uses a simple cost-estimation method to decide whether
producing a better plan is possible. The execution time of the
algorithm is exponential in the number of sub-queries in conjunction, therefore we have chosen
conjunction chains to benchmark the different implementations. Table \ref{fig:runtimes} shows the execution
times of the Query Optimizer.
The measurements were made with SICStus Prolog 3.12.5, on a machine
with an Intel\textregistered Pentium\textregistered M 2GHz Processor. The results show that the runtime
of the original algorithm is exponential, and that both the branch-and-bound and the memoization
techniques speed up the algorithm. However, memoization is not successful enough if the query has
many source predicates referring to the same source. This is because when dealing with source predicates,
the algorithm enumerates all the possible (callable) subsets of the source predicates in the query.
Memoizing cannot help in this situation, but branch-and-bound helps:
when using both the techniques, the size of the queries can be increased, no exponential increase in runtime can be observed.
The last row shows the optimization of a real-life query mustering up negation, aggregation,
disjunction, conjunction, constraint and source predicates.

\begin{table}[htbs]
\begin{center}
\begin{tabular}{|c|r|r|r|r|}
\hline
Query & Naive algorithm & Branch\&Bound & Memoizing & Both \\
\hline
\ 8 source preds, different source &   30.62   & 0.0961 &    0.1716  & \hspace{0.1cm}0.0128 \\
\ 9 source preds, different source &  276.74   & 0.2016 &    0.4053  & 0.0171 \\
10 source preds, different source  & 2766.28   & 0.2986 &    0.9584  & 0.0222\\
\ 8 source preds, common source    &  290.10   & 0.0696 &   15.2550  & 0.0141 \\
\ 9 source preds, common source    & 5298.83   & 0.1277 &  140.2700  & 0.0196 \\
10 source preds, common source     &    N/A    & 0.2071 & 1439.8800  & 0.0262 \\
specimen query                     &    0.0379 & 0.0250 &    0.0069  & 0.0082 \\
\hline
\end{tabular}
\caption{Execution times in seconds}
\label{fig:runtimes}
\end{center}
\end{table}

\section{Related Work}
\label{sec:related}

The compiler of Mercury, a pure declarative Prolog-variant does
predicate reordering according to the I/O modes of the
predicates, as described in \cite{somogyi96execution}.  The mode system of Mercury
is much more expressive than the mode system of SINTAGMA's Query
Optimizer, our \verb!in! and \verb!out! modes are easily handled by
the Mercury compiler. On the other hand, it does not offer optimizations similar to
our optimizer, it only reorders the predicates according to their I/O
modes.

The SIMS and the Infomaster information integration systems have a
query optimizer component, as described in \cite{hsu00semantic} and
\cite{duschka97query}, however, they have a different task than
ours. In those systems, query optimizers take advantage of semantic
knowledge about the information sources to choose a query plan that
needs the least number of information source accesses, among the plans which
answer the user query. In the Mediator of SINTAGMA, this is the task
of the Query Planner, and Query Optimizer optimizes only the query
execution plan.

\section{Conclusion}
\label{sec:conclusion}

In the SILK information integration system, query plans often needed
manual tuning, especially in the presence of information sources which
have I/O mode restrictions. While SILK had only conjunction and
disjunction in the query plans, queries in SINTAGMA contain also
negation and aggregation. With the growing use of information sources
other than relational databases, the need for manual tuning of the
more complex queries has become a major drawback. The
Query Optimizer presented in the article is a part of the next
release of SINTAGMA. During the testing of the system, some
details, especially the cost estimation formulas will be refined. With
the use of the Query Optimizer, we expect that manual tuning of query
plans will become unnecessary.

\section*{Acknowledgements}

The authors acknowledge the support of the Hungarian NKFP programme
for the SINTAGMA project under grant no.\ 2/052/2004. We would also
like to thank all the people participating in this project, but mostly
Tamás Benkő and Gergely Lukácsy.

\bibliographystyle{splncs}
\bibliography{QueryOptimizer.bib}

\begin{thebibliography}{1}

\bibitem{silk:iclp}
Benk\H{o}, T., Krauth, P., Szeredi, P.:
\newblock A {Logic}-based {System} for {Application} {Integration}.
\newblock In: {Proceedings of the International Conference on Logic
  Programming}. Volume 2401 of Lecture Notes in Computer Science., Springer
  (2002)  452--466

\bibitem{somogyi96execution}
Somogyi, Z., Henderson, F., Conway, T.:
\newblock The {Execution} {Algorithm} of {Mercury}, an {Efficient} {Purely}
  {Declarative} {Logic} {Programming} {Language}.
\newblock Journal of Logic Programming \textbf{29} (1996)  17--64

\bibitem{DBLP:conf/acm/WangYC93}
Wang, J., Yoo, J.P., Cheatham, T.J.:
\newblock Efficient {Reordering} of {C-PROLOG}.
\newblock In: ACM Conference on Computer Science. (1993)  151--155

\bibitem{ioannidis96query}
Ioannidis, Y.E.:
\newblock Query optimization.
\newblock ACM Computing Surveys \textbf{28} (1996)  121--123

\bibitem{chaudhuri98overview}
Chaudhuri, S.:
\newblock An overview of query optimization in relational systems.
\newblock (1998)  34--43

\bibitem{chaudhuri95overview}
Chaudhuri, S., Shim, K.:
\newblock An {Overview} of {Cost-based} {Optimization} of {Queries} with
  {Aggregates}.
\newblock Data Engineering Bulletin \textbf{18} (1995)  3--9

\bibitem{hsu00semantic}
Hsu, C.N., Knoblock, C.A.:
\newblock Semantic {Query} {Optimization} for {Query} {Plans} of
  {Heterogeneous} {Multidatabase} {Systems}.
\newblock Knowledge and Data Engineering \textbf{12} (2000)  959--978

\bibitem{duschka97query}
Duschka, O.M.:
\newblock Query Planning and Optimization in Information Integration.
\newblock PhD thesis (1997)

\end{thebibliography}

\end{document}